\documentclass[a4paper]{jpconf}
\usepackage{graphicx}

\usepackage{lineno}

\begin{document}
\title{Novel techniques of imaging interferometry analysis to study gas and plasma density for laser-plasma experiments}

\author{F. Filippi, M. Cipriani, S. Mastrostefano, M. Scisciò, F. Consoli}

\address{ENEA, Nuclear Department, Centro Ricerche Frascati, 00044, Frascati, Roma, Italia}

\ead{francesco.filippi@enea.it}

\begin{abstract}
Laser-plasma based experiments are always more demanding about the plasma features which need to be generated during the interaction. This is valid for laser-plasma acceleration as well as for inertial confinement fusion experiments. Most of these experiments are moving toward high repetition rate operation regimes, making even more demanding the requests on the plasma sources and the diagnostics to be implemented.\\
Interferometry is one of the most used methods to characterize these sources, since it allows for non-perturbative, single-shot measurements either of the neutral gas or the plasma density. The design of the interferometric setup is non-trivial and needs to be shaped on the actual conditions of the experiment. Similarly, the analysis of the raw data is a complex task, prone to many sources of error and dependent on the manual inputs.\\
In this work, we will present the techniques we are developing for the analysis of the interferograms to measure both the gas and plasma density. We will show the methods, the progress and the problems we encountered in the development of novel routines of analysis based on machine learning. The architectures and the methods to obtain data used for training and testing them will be introduced. The study is ongoing and preliminary results with synthetic data will be presented. The goal is to set up a fast and operator independent diagnostic for the feedback of plasma sources toward high repetition rate experiments.

\end{abstract}

\section{Motivation}

Through the interference among two coherent laser beams, namely the reference and the probe beams, it is possible to measure the phase shift experienced by the probe caused by the variation of the refractive index due to the presence of plasma free-electrons or a non-ionized gas. The phase shift can be expressed by the following formula 
\begin{equation}\label{eq:interferometry}
\Delta \phi = \frac{2\pi}{\lambda}\int_{L} 1-N_\lambda (z) dz
\end{equation}
where $L$ is the length of the path that the laser has crossed through the plasma, $z$ is the propagation axis along the probe, $\lambda$ is the laser wavelength and $N_\lambda$ is the refractive index of the sample, meanwhile the refractive index encountered by the reference beam is assumed to be 1. $N_\lambda$ depends on the type of sample analyzed. In typical laser-plasma experiments this may be a non-ionized gas, in this case $N_\lambda$ is function of the pressure, temperature and polarizability of the gas molecule, or a plasma, in which $N_\lambda$ depends on the plasma free-electron density and on the probe laser wavelength.\\
In imaging interferometry, the probe and the reference beams, usually originated by splitting the same laser beam, are recombined on a specific plane to create an interference pattern, called \textit{interferogram}. The phase shift is made evident by the fringe deflection, which highlights the area where the different refractive index causes a phase shift in the probe. This method permits to have a quantitative measurement of the plasma electron/gas density and its spatial distribution. Still, the analysis of the raw data is a non-trivial task, prone to many sources of error and dependent on manual inputs whose consequences we tried to mitigate in other works \cite{Filippi2023}. In future laser-plasma experiments the repetition rate of the laser shots will be dramatically increased, creating a severe constraint on the imaging diagnostics and their analysis. High repetition rate requires tools able to furnish analyzed data in real-time with high affordability. At the present days, most of the used routines, based on the Fast Fourier Transform and Abel or other inversion algorithms \cite{Nugent1985, Tomassini2001}, require an action from the operator and even minor changes (e.g. the region of interest of the image) require the operator to adjust the analysis parameters, resulting in time loss and potentially causing errors. This is particularly important for the characterization of laser-produced plasma density, and it is crucial for the tuning of the laser and target properties to reach the optimal laser-plasma interaction in inertial confinement fusion, high energy density research and laser-driven plasma wakefields accelerators.

In this article, we will show the progress in developing an automated tool for a fast and operator-independent analysis of the interferograms \cite{Filippi2025}. This could enable a real time feedback to help the operation of laser-plasma at high repetition rate. In this framework, we are developing a machine learning (ML) routine, which has been trained to fast analyse the interferograms and retrieve the corresponding phase shift. We will introduce the architecture we decided to use, the cGAN structure, and the methods we used to train the ML network. This architecture has been already used for the translation of aerial images into maps, or the colorization of pictures \cite{Isola}. Similar approaches have been developed for scientific data analysis (e.g. \cite{Lee2020, Rivenson2019, Ronneberger2015}). The program we have developed so far is able to retrieve the phase shift of a synthetic interferogram. We will discuss the further steps to improve it for the analysis of raw interferograms, acquired in real experimental conditions. This is the first step toward the implementation of an automated tool to retrieve information from an interferogram in a fast and robust way for high repetition rate experiments. 

\section{Methods: Network and Dataset}
\begin{figure}
	\centering
	\includegraphics[width=0.8\linewidth]{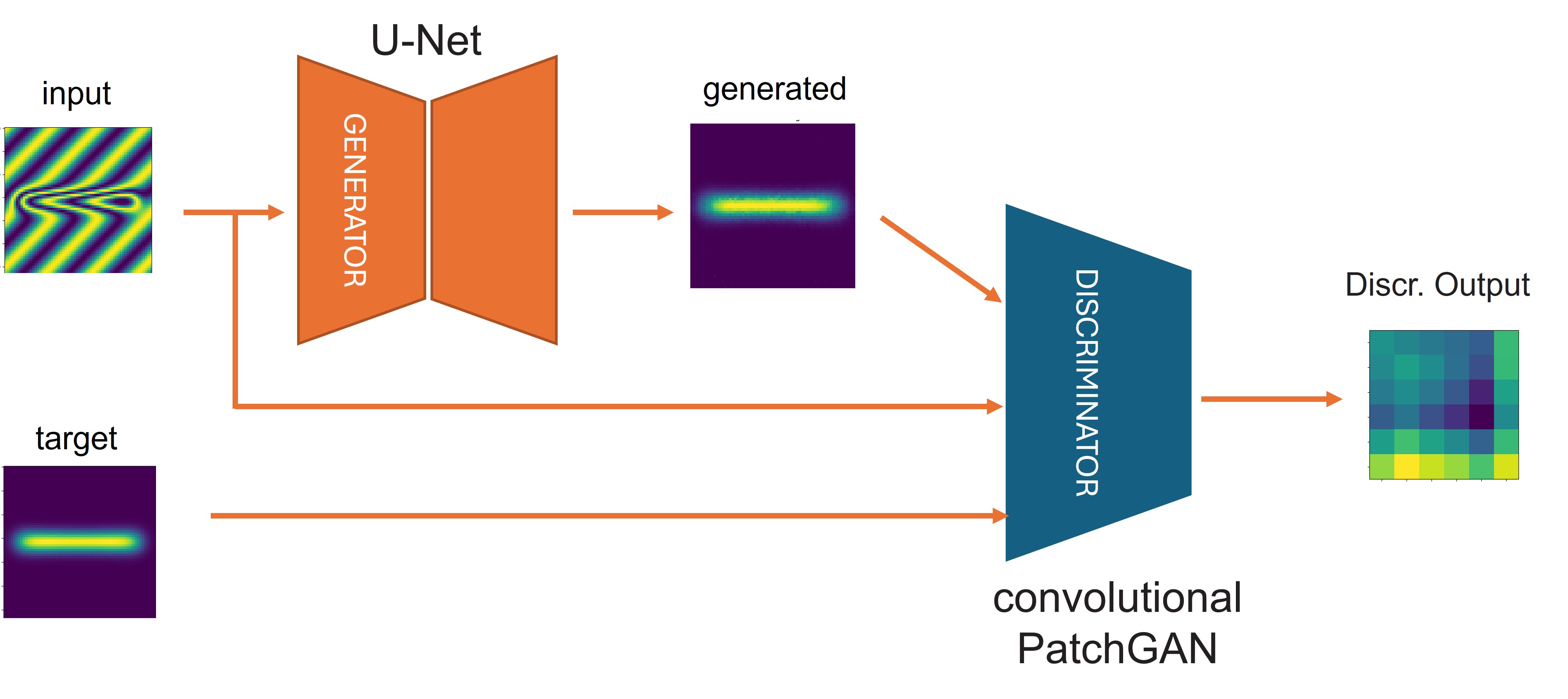} 
	\caption{Schematic of the cGAN architecture and the training workflow. During the "training", the input images are sent to the generator which produces output images. The discriminator received, either, a couple of input and output images, or a couple of input and target images. During the training, the discriminator is optimized to recognise the outputs as "false" and the targets as "true", while the discriminator is optimized to produce images the most similar possible to the targets. The optimization parameters are given by the "loss functions".}
	\label{fig:Schematic}
\end{figure}

Machine learning provides fast networks which map the inputs to the outputs through a process called \textit{training}. In the conditional Generative Adversarial Network (cGAN)\cite{Isola} two models are trained simultaneously by an adversarial process. A network called \textit{generator} learns to create an image starting from an input image. A different network, called \textit{discriminator}, learns to recognize real images, associated to the input image, and distinguish them from "fakes", actually produced by the generator (as shown in Fig. \ref{fig:Schematic}). During the \textit{training}, the generator progressively becomes better in producing images similar to the expected outputs. Meanwhile, discriminator becomes better in distinguish the original outputs from the ones generated by the generator. Training is guided by \textit{loss functions} that balance the ability of both the networks. Details on this function can be found in \cite{Isola}. The process repeats those steps iteratively until it reaches an equilibrium when the discriminator can no longer distinguish real images from fakes. 

The generator we used is based on a "U-Net" architecture \cite{Ronneberger2015}. The generator has to map a high resolution input image to a high resolution output image. The input is passed through a series of layers that progressively downsample until a bottleneck layer, from which the process is reversed and the output image can be progressively reconstructed through a series of similar layers. This process may cause a loss of "low-level" information which should be better to transfer from the input to output. Then, a series of shortcuts let the information to skip some connections and to pass through corresponding layers that have the same sized feature maps. 

The discriminator uses a convolutional "PatchGAN" classifier \cite{Isola}. The discriminator takes the input image and an image which may be produced by the generator or be the real image expected to be associated to the input and predicts whether the second image is real or not. Since the training process can be tuned to be particularly efficient for the low-spatial-frequency features, this particular architecture is optimized to be sensitive to high-spatial-frequency variation and penalizes structure at the scale of portion of the image (patches) whose dimension depends on the number of layers of the discriminator. This also let to reduce the dimension of the discriminator i.e. the parameters to train, making the training faster. Still, it can be applied to arbitrarily large images. The output of the discriminator is a map of the "trust" of the discriminator on each patch of the input couple of images. This is used during the training for the optimization of the networks and for the calculation of the loss function.

To train and implement those networks, we used the open-source library TensorFlow, a specific tool developed for machine learning and artificial intelligence. This library can be handled through Python, an open source programming language, and the interface Keras \cite{Chollet}. Different communities have successfully applied this approach to a large variety of image-to-image translation tasks, not only for scientific purposes \cite{Isola,Lee2020,Rivenson2019,Ronneberger2015}. The high adaptability, the large community working on it and the possibility to retrieve information online is a significant advantage of this architecture.

\begin{figure}
	\centering
	\includegraphics[width=0.8\linewidth]{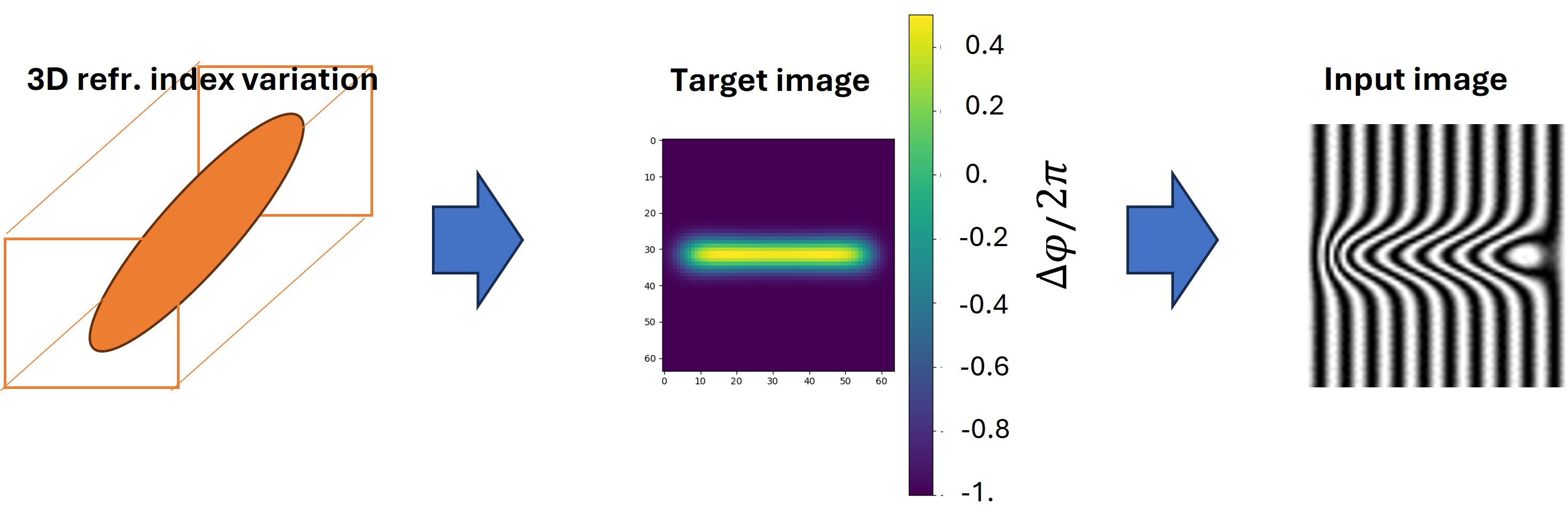} 
	\caption{Synthetic data generation workflow.}
	\label{fig:Syntetic}
\end{figure}

To train the network, two sets of input data (\textit{dataset}) are needed: \textit{inputs} and \textit{targets} (or ground truth). In our case, the inputs correspond to the interferograms, while targets are the map of the phase shift relative to the corresponding interferogram. We used couples of synthetic interferograms/phase shifts to train the network instead of real interferograms coupled with the phase shift calculated with an already-developed routine. This to ensure that even small errors in the analysis do not interfere with the training. Images have been numerically generated from synthetic 3D refractive index distribution as schematically described in Fig. \ref{fig:Syntetic}. Longitudinally, we varied the length of the plasma and the gradient at its ends with a hyperbolic tangent function. The transverse profile has been modeled with the Laguerre-Gaussian modes. The width of the distribution has been varied longitudinally, to better reproduce scenarios similar to experimentally acquired data. This means that the distribution may be wider at one end, and narrower at the other end.
Through the equation \ref{eq:interferometry}, we can evaluate the integrated phase shift of a probe laser crossing the distribution transversely. We reconstructed the interferograms by converting the phase shift in fringe shift. Per each phase shift image, we obtained several interferograms varying the fringe spatial frequency and their orientation.

During the training, the routine which loads the data randomly rotate, crop, add noise and change the intensity of the input images, re-adapting the target images to the new corresponding input. This makes the net more robust and able to face the “real” data, mitigating the effects of the overfitting \cite{Chollet}.

\section{Training and preliminary results}
\begin{figure}
	\centering
	\includegraphics[width=0.7\linewidth]{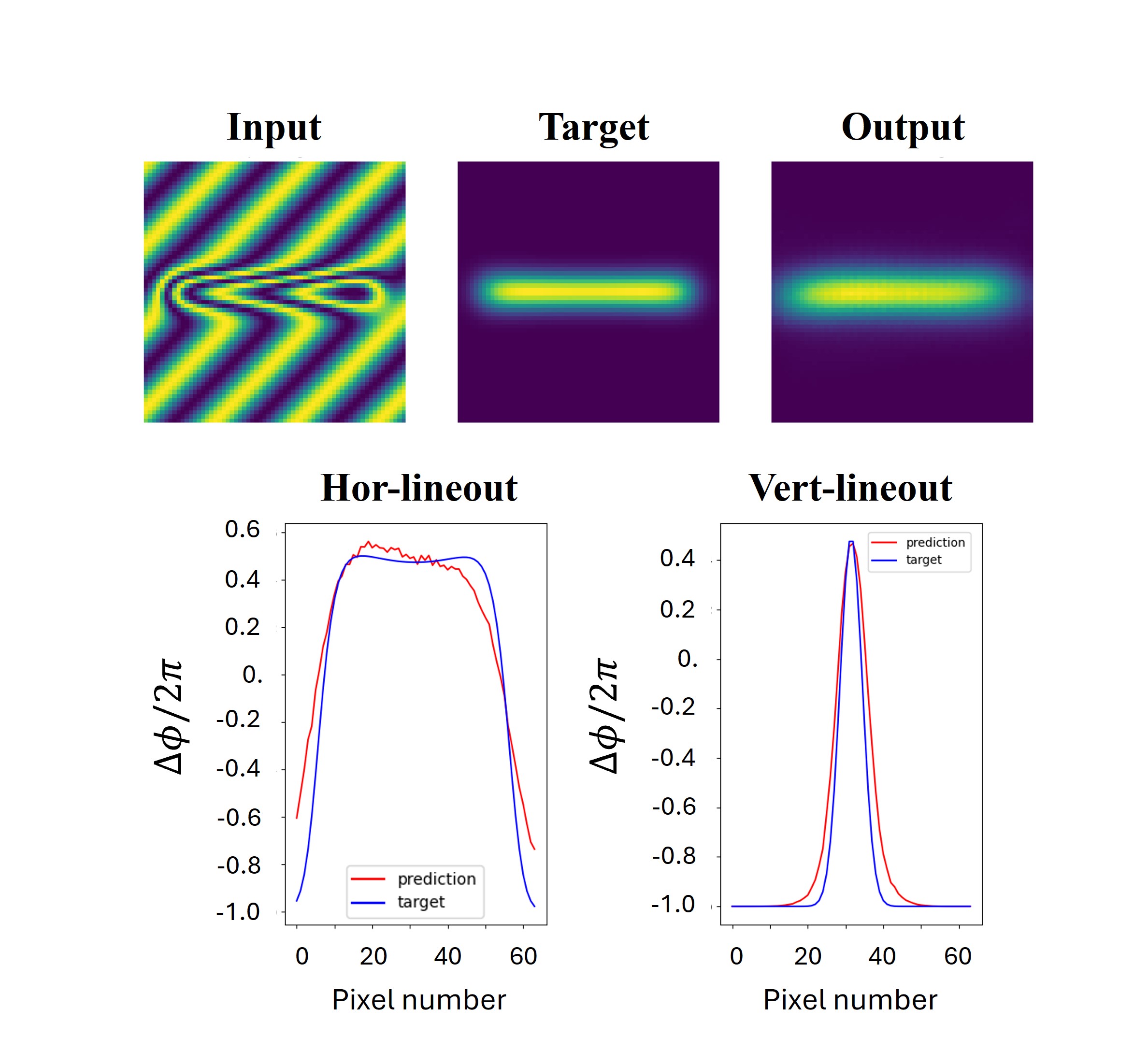} 
	\caption{Input image and target image generated from synthetic test data, with the corresponding output image predicted by the generator. In the bottom line there are the vertical and horizontal lineouts of the central line and column of the generated image (in red) and of the target image (in blue). }
	\label{fig:TrainResults}
\end{figure}

We trained the cGAN over more than $4400$ input-target data. 
At the end of the training, we tested the network on a set of interferograms not used for the training. In Fig.\ref{fig:TrainResults} there is an example of one of those tests. In the top left of the figure, there are the input image, i.e. the synthetic interferogram. The phase shift from which the interferogram is generated is shown in the top centre of the figure. The output phase shift generated by the generator is in the top-right. A comparison between the lineouts of the two phase shift images is present in the bottom line of the figure. The retrieved phase shift (normalized to $2\pi$) is quite noisy, but its variation is low compared to the maximum value of the dephasing. The lineouts of the target (blue) and generated (red) are shown in the bottom of the figure. As can be seen, the discrepancy with the original data is not particularly pronounced in the central region of the horizontal lineout, while it is more evident at the edges where slope of the phase shift is larger and goes rapidly to zero. The vertical lineout shows a good agreement between the maximum value of the target data and the one obtained from the trained generator. The output image correctly goes to zero, still having a small discrepancy at the lower values. This suggest that the network is less accurate when steeper slopes of density are analysed. Still, the generated phase shift well reproduce the shape and the maximum value of the target image and the effects of those discrepancies on the density measurement will be analysed in the next steps of this work.

We used the same network to analyze a 'real' interferogram acquired in an experimental campaign. In this test, the retrieved phase shift was lower than the correct value. The variation of the fringes are correctly recognised by the network, but the entire shape of the phase shift is not correctly reconstructed. This issue will be addressed in the follow up of our study.

\section{Conclusions and future works}

We investigated the possibility to implement fully-automated and operator-independent interferogram analysis for gas and plasma density measurements. We found in the cGAN an appropriate structure for the image-to-image translation to retrieve the phase shift from an interferogram. We developed a routine for the generation of synthetic interferograms for training and testing the net. We achieved good results with synthetic images. Indeed, the trained network could reconstruct the phase shift with good accuracy. We did not reach the same results with experimentally acquired interferograms whose phase shift was not correctly reconstructred.  

In the following, we will use experimental images to train the network, hopefully reaching a major stability in analysing different input images. We are also implementing an interferometry/shadowgraphy setup to acquire at the same time with the same optical system the interferograms and the shadowgrams of a target in typical fusionistic experiments. We aim to use those images to increase the reliability of the software. Finally, we will implement a 3D reconstruction of the plasma density to provide a quick, reliable reconstruction of the sample analyzed.

\section{Acknowledgment}
This work has been carried out within the framework of the EUROfusion Consortium, funded by the European Union via the Euratom Research and Training Programme (Grant Agreement No. 101052200–EUROfusion). Views and opinions expressed are however those of the author(s) only and do not necessarily reflect those of the European Union or the European Commission. Neither the European Union nor the European Commission can be held responsible for them.


\section*{References}
\begin{thebibliography}{9}
\bibitem{Filippi2023} Filippi F., et al. \textit{Plasma density profile reconstruction of a gas cell for Ionization Induced Laser Wakefield Acceleration}. Journal of Instrumentation 18.05 (2023): C05013.
\bibitem{Nugent1985} K.A.Nugent, \textit{Interferogram analysis using an accurate fully automatic algorithm}. Applied Optics 18, 3101 (1985).
\bibitem{Tomassini2001} Tomassini P, et al. \textit{Analyzing laser plasma interferograms with a continuous wavelet transform ridge extraction technique: the method}. Applied Optics 2001; 40(35):6561–8. 
\bibitem{Filippi2025} Filippi F., et al. \textit{Toward an automated tool for interferogram analysis for real time characterization of plasma density profile in laser produced plasmas}. Accepted for publication.
\bibitem{Isola} Isola P. et al ., {\it Image to Image Translation with Conditional Adversarial Networks}. ArXive (2017). DOI: 10.48550/arXiv.1611.07004
\bibitem{Lee2020} Lee S. et al., \textit{Deep learning for high-resolution and high-sensitivity interferometric phase contrast imaging}. Sci Rep. 2020;10(1):1–12. 
\bibitem{Rivenson2019} Rivenson Y, Wu Y, Ozcan A., \textit{Deep learning in holography and coherent imaging}. Light Sci Appl. 2019;8(1). 
\bibitem{Ronneberger2015} Ronneberger O., et al. \textit{U-net: Convolutional networks for biomedical image segmentation." Medical image computing and computer-assisted intervention}. MICCAI 2015: 18th international conference, Munich, Germany, October 5-9, 2015.
\bibitem{Chollet} Chollet, F., {\it Deep learning with Python}. Simon and Schuster, 2021.

\end{thebibliography}
\end{document}